\documentclass[a4paper,floatfix,rmp,twocolumn,showkeys,superscriptaddress]{revtex4}
	\usepackage[latin1]{inputenc}
	\usepackage{graphicx}
	\usepackage{listings}
	\usepackage{lipsum}

\begin{document}

\title{The morphospace of language networks}

\providecommand{\MIT}{Department of Physics, Massachusetts Institute of Technology, USA}
\providecommand{\ICREA}{ICREA-Complex Systems  Lab, Universitat Pompeu Fabra,  08003   Barcelona}
\providecommand{\IBE}{Institut de Biologia Evolutiva (CSIC-UPF), 08003 Barcelona}
\providecommand{\SFI}{Santa Fe Institute, 399 Hyde Park Road, Santa Fe NM 87501, USA}

\author{Lu\'is F Seoane} \affiliation{\MIT} 
\author{Ricard Sol\'e} \affiliation{\ICREA} \affiliation{\IBE}  \affiliation{\SFI} 

	\vspace{0.4 cm}

	\begin{abstract}
		\vspace{0.2 cm}

		Language can be described as a network of interacting objects with different qualitative properties and complexity. These networks include semantic, syntactic, or phonological levels and have been found to provide a new picture of language complexity and its evolution. A general approach considers language from an information theory perspective that incorporates a speaker, a hearer, and a noisy channel. The later is often encoded in a matrix connecting the signals used for communication with meanings to be found in the real world. Most studies of language evolution deal in a way or another with such theoretical contraption and explore the outcome of diverse forms of selection on the communication matrix that somewhat optimizes communication. This framework naturally introduces networks mediating the communicating agents, but no systematic analysis of the underlying landscape of possible language graphs has been developed. Here we present a detailed analysis of network properties on a generic model of a communication code, which reveals a rather complex and heterogeneous morphospace of language networks. Additionally, we use curated data of English words to locate and evaluate real languages within this language morphospace. Our findings indicate a surprisingly simple structure in human language unless particles are introduced in the vocabulary, with the ability of naming any other concept. These results refine and for the first time complement with empirical data a lasting theoretical tradition around the framework of {\em least effort language}. 

	\end{abstract}


	\maketitle

	\section{Introduction}
		\label{sec:1}
		
    The origins of complex forms of communication, and of human language in particular, defines one of the most difficult problems for evolutionary biology \cite{Bickerton1992, SzathmaryMaynardSmith1997, Deacon1998, Bickerton2014, BerwickChomsky2015}. Language makes our species a singular one, equipped with an extraordinary means of transferring and creating a virtually infinite repertoire of sentences. Such an achievement represents a major leap over genetic information and is a crucial component  of our success as a species \cite{Suddendorf2013}. Language is a specially remarkable outcome of the evolution of cognitive complexity \cite{JablonkaLamb2006, JablonkaSzathmary1995} since it requires perceiving the external world in terms of objects and actions and name them using a set of signals. 

		Modelling language evolution is a challenging issue, given the unavoidable complexity of the problem and its multiple facets. Language evolution takes place in a given context involving ecological, genetic, cognitive, and cultural components. Moreover, language cannot be described as a separate collection of phonological, lexical, semantic, and syntactic features. All of them can be relevant and interact with each other. A fundamental issue of these studies has to do with language evolution and how to define a proper representation of language as an evolvable replicator \cite{ChristiansenCulicover2016}. Despite the obvious complexities and diverse potential strategies to tackle this problem, a common feature is shared by most modelling approximations: an underlying bipartite relationship between signals (words) used to refer to a set of object, concepts, or actions (meanings) that define the external world. Such mapping asumes the existence of speakers and listeners, and is used in models grounded in formal language theory \cite{Nowak2002}, evolutionary game theory \cite{NowakKrakauer1999}, agent modelling \cite{Steels1997,Kirby2001, Kirby2002,KirbySmith2008,Steels2015}, and connectionist systems \cite{CangelosiParisi1998}. 

		\begin{figure*}[t]
			\begin{center}
				\includegraphics[width = 0.75 \textwidth]{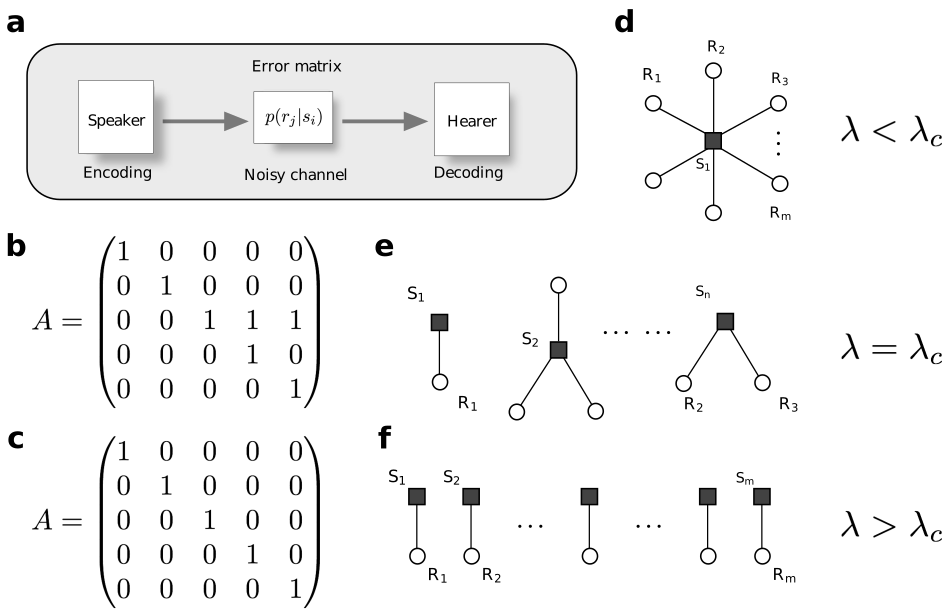}

        \caption{ {\bf A toy model to explore {\em least effort language}. } {\bf a} Any minimal model of communication should include a (possibly noisy) channel that connects hearers and speakers. At the heart of this channel lies a confusion matrix $p(r_j | s_i)$ that tells the likelihood that an object is interpreted by the hearer when a signal is uttered by the speaker. In an ideal, not noisy situation we can encode these word-object associations by a matrix ({\bf b} and {\bf c}) such that $a_{ij} = 1$ if signal $s_i$ names object $r_j$ and $a_{ij} = 0$ otherwise. Such matrices naturally introduce synonymy and polysemy. They also define bipartite {\em language networks} ({\bf d-f}). We study how an optimization problem posed on the communication channel is reflected in optimal languages, with extreme solutions resulting in minimal effort for a speaker (hence maximal for a hearer, {\bf d}) or the other way around ({\bf f}). }
				\label{fig:01}
			\end{center}
		\end{figure*}

		In all these approaches, a fundamental formal model of language includes (figure \ref{fig:01}{\bf a}): i) a speaker that encodes the message, ii) a hearer that must decode it, and iii) a potentially noisy communication channel \cite{CoverThomas1991} described by a set of probabilities of delivering the right output for a given signal. Within the theory of communication channels, key concepts such as reliability, optimality, or redundancy are of high relevance to the evolution of language. 

		In looking for universal rules pervading the architecture and evolution of communication systems, it is essential to consider models capable of capturing the very basic properties of language. Such a minimal toy model \cite{FerrerSole2003} can be described as a set
			\begin{eqnarray}    
				S &=& \{s_i, i=1, \dots, n\}
				\label{eq:01}
			\end{eqnarray}   
		of available signals or ``words'', each of which might or might not name one element from the set 
			\begin{eqnarray}
				R &=& \{ r_j, j=1, \dots, m\}
			\end{eqnarray}
		of objects or ``meanings'' existing in the world. These potential associations can be encoded by a matrix $A \equiv \{a_{ij}\}$ such that $a_{ij}=1$ if signal $s_i$ names object $r_j$ and $a_{ij}=0$ otherwise (figure \ref{fig:01}{\bf e-f}). 
		
		Following a conjecture made by George Zipf \cite{Zipf1949}, this model was used to test whether human language properties could be the result of a simultaneous minimization of efforts between hearers and speakers \cite{FerrerSole2003}. In a nutshell, if a signal in language $A$ can name several objects in $R$, its degeneracy implies a large decoding effort $\Omega_h$ to the hearer. A limit case is shown in figure \ref{fig:01}{\bf d}, where one signal names all objects. Otherwise, if one (and only one) different signal exists to name each of the elements in $R$ (figure \ref{fig:01}{\bf c} and {\bf f}), the burden $\Omega_s$ falls mainly on the speaker who must find each precise name among all those existing, while a hearer does not incur in any decoding costs. Minimal effort for one of the parts implies maximal cost for the other. Zipf's conjecture suggested that  a compromise between these two extremes would pervade the efficiency of human language. 

		The model introduced above \cite{FerrerSole2003} allows us to quantify these costs explicitly and hence tackle the Zipfian {\em least effort principle} using information theory. It does so by considering a linear `energy' function $\Omega(\lambda)$ that optimal languages would minimize, and that contains both the hearer and speaker costs:
      \begin{equation}
        \Omega(\lambda) = \lambda \Omega_h + (1-\lambda) \Omega_s. 
        \label{eq:02}
      \end{equation}
		$\lambda \in [0,1]$ is an external metaparameter balancing the importance of both contributions. In terms of information theory, it is natural to encode $\Omega_s$ and $\Omega_h$ as entropies. One choice is to define $\Omega_h$ as the conditional entropy that weights the errors made by the hearer, namely:
			\begin{eqnarray}			 
        H_m(R|s_i) &=& -\sum_{j=1}^m p(r_j|s_i) log_m p(r_j|s_i), \\ 
        H_m(R|S) &=&  \sum_{i=1}^n p(s_i) H_m(R|s_i) \equiv \Omega_h; 
				\label{eq:03}
			\end{eqnarray}
		where $p(r_j | s_i)$ is the probability that object $r_j$ was referred to when the word $s_i$ was uttered by a speaker. Such {\em confusing} probabilities depend on the ambiguity of the signals. We can also postulate the following effort for a speaker: 
			\begin{eqnarray}
        H_n(S) &=& -\sum_{i=1}^n p(s_i) log_n(p(s_i)) \equiv \Omega_s, 
				\label{eq:04}
			\end{eqnarray}
		where $p(s_i)$ is the frequency with which the $s_i$ signal is employed given the matrix $A$. To compute $p(s_i)$ we assume that every object needs to be recalled equally often and that we choose indistinctly among synonyms for each object. 

		The global minimization of equation \ref{eq:02} was tackled numerically \cite{FerrerSole2003} and analytically \cite{ProkopenkoPolani2010, SalgeProkopenko2013}. Slight variants of the global energy have also been studied, broadly reaching similar conclusions. An interesting finding is the presence of two ``phases'' associated to the extreme solutions shown in figures \ref{fig:01}{\bf d} and {\bf f}. These two regimes were associated to rough representations of a ``no-communication possible'' scenario in which one signal can name any object (figure \ref{fig:01}{\bf d}) and a phase tied to animal and computer programming languages where non-ambiguous (one-to-one) mappings would be found (figure \ref{fig:01}{\bf f}). The two phases recover the ideal solutions for speakers and hearers respectively, and they are separated by an abrupt transition at a given critical value $\lambda_c$. It was conjectured that human language would exist right at this critical point. 

		Solutions of this linear global optimization problem have been found \cite{ProkopenkoPolani2010, SalgeProkopenko2013} and they display a mixture of properties, some associated (and some others not) to human language features. There might be a potential limitation with this approach: is the linear constraint a reasonable assumption? If no predefined coupling between $\Omega_h$ and $\Omega_s$ is introduced, the simultaneous optimization of both {\em targets} becomes a {\em Multi Objective} (or Pareto) {\em Optimization} (MOO) problem \cite{Deb2003,Coello2006, Schuster2012, Seoane2016}. This is a much more general approach that does not make additional assumptions about the existence of a global energy such as equation \ref{eq:02}. The solutions to MOO problems are not a single global optimum, but a collection of designs (in this case, word-object associations encoded by matrices) that constitute the optimal tradeoff between our optimization targets. This tradeoff (called the Pareto front) and its shape have recently been linked to thermodynamics, phase transitions, and critical phenomena \cite{SeoaneSole2013, SeoaneSole2015a, SeoaneSole2015b, SeoaneSole2015c, Seoane2016}. By relaxing the assumptions concerning the energy function, a more general scenario is considered. 

		The Pareto front for the MOO of language networks has never been portrayed. In this paper we aim at fully exploring the space of communication networks in the speaker/hearer effort space where the Pareto front defines one of its boundaries. It will be shown that the front matches the global minimization problem only at the critical point. But we will also study the whole space of language networks beyond the Pareto front, showing that it exists a wealth of communication codes embodied by all different binary matrices. These, as they link signals and objects, naturally define graphs with important information about how easy communication is, how words relate to each other, or how objects become linked in semantic webs as a same signal refers to many of them. All these characteristics pose interesting, alternative driving forces that may be optimized near the Pareto front or, in the contrary, might pull actual communication systems away from it. 

		By exploring the whole space of possibilities we are defining a {\em morphospace} of language networks. The concept of {\em theoretical morphospace} \cite{McGhee1999} was introduced within evolutionary biology \cite{Raup1965,Niklas1997,Niklas2004} as a systematic way of exploring all possible structures allowed to occur in a given system. This includes real (morphological) structures as well as those resulting from theoretical or computational models. Typically the morphospace is constructed in one of two different ways. One is applied to real sets of data. In this case, available morphological traits defined on each system are measured and a statistical clustering method (such as principal component analysis) is applied as a way to define the main axes and locate each system within this space \cite{McGhee1999}. The alternative is to use explicit parameters that define continuous axes that allow ordering all systems in a properly defined metric space. In recent years, graph morphospaces have been explored, thus showing how the concept can be generalized to the analysis of complex networks \cite{Avena2014}. In our context, the language morphospace analyzed below is shown to be unexpectedly rich. It appears partitioned into a finite set of language networks, thus suggesting archetypal classes involving distinct type of communication graphs. This also occurs within the set of optimal communication networks that define the Pareto front of the morphospace. 

		Finally, dedicated, data-driven studies exist about different optimality aspects of language, from prosody to syntax among many others \cite{JaegerLevy2006, FrankJaeger2008, Jaeger2010, PiantadosiGibson2011, MahowaldGibson2013}. But discussion of the least-effort language model has focused on its information theoretical characterization. The hypothesis that human language falls near the phase transition of the model has never been tested on empirical data before. We do so here using the WordNet database \cite{Fellbaum1998, Miller1995}. The previous development of the morphospace allows us not only to asses the optimality of real corpora, but also to portray some of its complex characteristics. This kind of study may become relevant for future evolutionary studies of communication systems, most of them relying on the ``speaker to noisy-channel to hearer'' scheme (figure \ref{fig:01}) at the core of the least effort model.

	\section{Complexity of language morphospace}	
		\label{sec:3}

		\begin{figure*}[t]
			\begin{center}
				\includegraphics[width = 0.75 \textwidth]{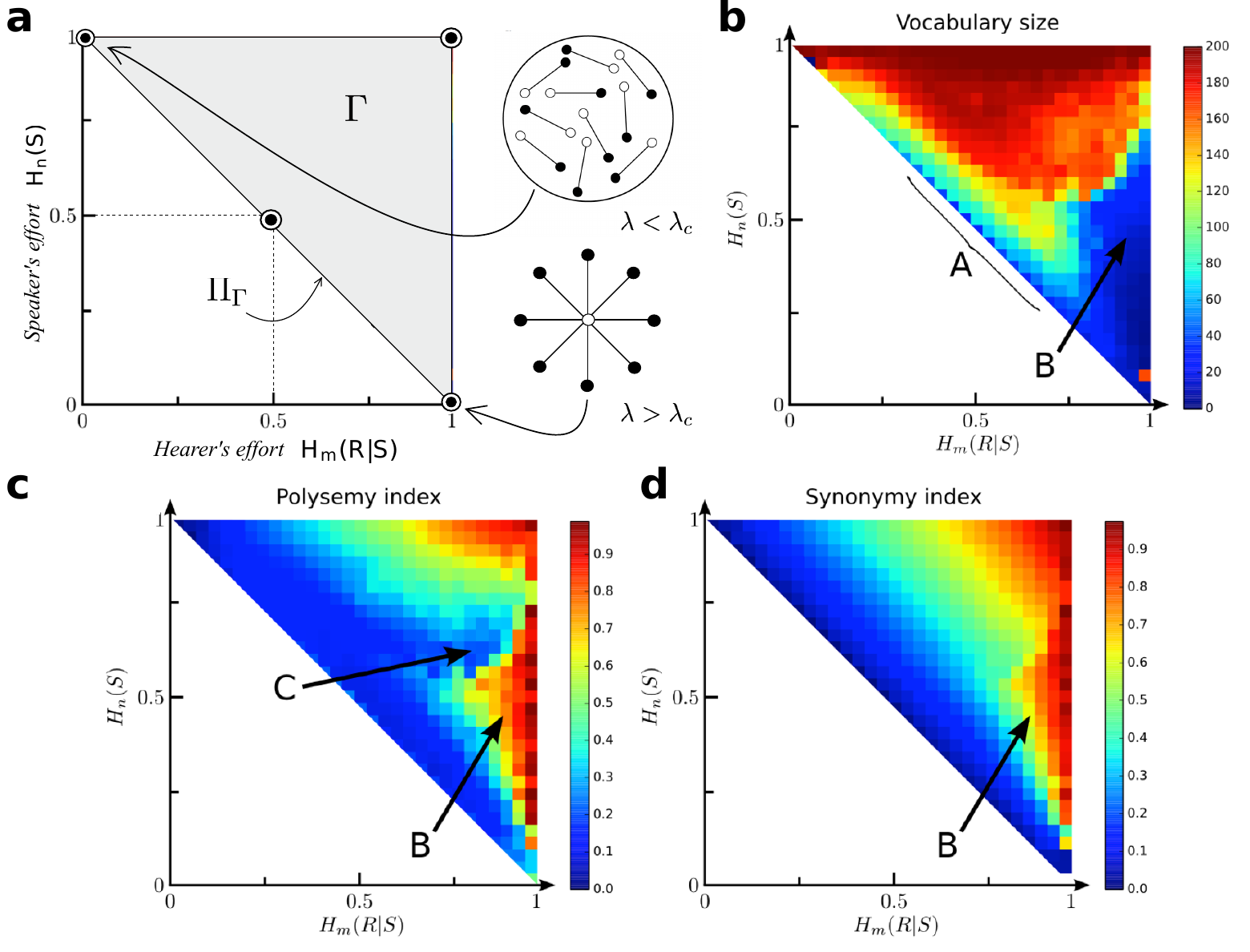}

        \caption{{\bf Vocabulary size, polysemy, and synonymy across language morphospace.} {\bf a} The space $\Gamma$ that can be occupied by language networks is shown in gray. Two limit cases (the one-to-one and star graphs) are also mapped. {\bf b} Effective vocabulary size is only low near the star graph (in a prominent area labeled B) and along the Pareto front. {\bf c} Polysemy is large in region B and as we complete the matrix $A$ towards the upper-right corner. {\bf d} Synonymy increases uniformly as we move apart from the front except for codes in B. This makes them highly Pareto inefficient. }
				\label{fig:02}
			\end{center}
		\end{figure*}

		In this section we characterize the morphospace of all codes allowed by our toy model. We refer to this set of possible languages as $\Gamma$. We look at it with the two least-effort target functions ($\Omega_h \equiv H_m(R|S)$ and $\Omega_s \equiv H_n(S)$) as a reference. Therefore, first it was necessary to find its boundaries in the $\Omega_h-\Omega_s$ plane, and to generate a fair sample throughout. In appendix \ref{app:01} we discuss thoroughly how this was done. Figure \ref{fig:02} shows the boundaries found for our morphospace, as well as the location of some prominent solutions: i) the star graph, which minimizes the effort of a speaker and maximizes that of a hearer; ii) the one-to-one mapping, often associated to animal communication, which minimizes the effort of a hearer at the expense of a speaker's; and iii) the Pareto optimal manifold ($\Pi_\Gamma$) corresponding to the lower, diagonal boundary of $\Gamma$ in the $\Omega_h-\Omega_s$ plane. $\Pi_\Gamma$ tells us the optimal trade-off between both targets. In appendix \ref{app:01} we discuss how the shape of this Pareto front implies that the model indeed has a first order phase transition, and determines analytically that this transition does contain a critical point. See \cite{SeoaneSole2013, SeoaneSole2015a, SeoaneSole2015b, SeoaneSole2015c, Seoane2016} for thorough discussions of the connection between the geometry of the Pareto front, phase transitions, and criticality. The criticality of this model had been suggested, but never proved analytically. Our results from appendix \ref{app:01} and the analytical findings of \cite{ProkopenkoPolani2010, SalgeProkopenko2013} imply that the Pareto front consists of all languages without synonyms. 

		To explore the morphosapce we take a series of measurements upon the $A$ matrices that relate to their size, network structure, or suitability as a model of actual human language. Far from smooth, simple gradients, we find a fragmented morphospace where different properties peak or fade, and in which languages are non-trivially clustered together. Such measurements where taken both on samples of languages across the morphospace and in the most restricted Pareto front. In the following, we report results for the morphospace in general. All results for the Pareto front alone can be found in appendix \ref{app:02}.

		\subsection{Characterizing the vocabulary}
			\label{sec:3.1}

			First of all we measure the effective vocabulary size (number of signals that refer to at least one object) of the codes ($L$, figure \ref{fig:02}{\bf b}), which can go from $L=1$ for the star graph to the one-to-one map ($L=n$). By plotting $L$ across the morphospace a non-trivial structure is revealed. Codes with small $L$ occur mostly near the star and in a narrow region adjacent to the Pareto front (marked A in figure \ref{fig:02}{\bf b}). Far apart from the front there is yet another region (marked B) with less than $30\%$ of all available signals being used. The transition to codes that use more than $75\%$ of available signals (central, red region in figure \ref{fig:02}{\bf b}) seems to be abrupt wherever we approach those codes from.

      It is important to take into account the effective vocabulary size when measuring certain properties. Let us consider a polysemy index $I_P$ and synonymy index $I_P$, defined as:
				\begin{eqnarray}
          I_{P} &=& \sum_{s_i \in S} {log_m(\rho_i) \over L}, \nonumber \\
          I_{S} &=& \sum_{r_j \in R} {log_L(\sigma_j) \over m}; 
					\label{eq:07}
				\end{eqnarray}
			respectively. Here $\sigma_j$ is the number of signals associated to object $r_j$ and $\rho_i$ is the number of objects associated to signal $s_i$. These indexes measure the average logarithm of $\sigma_j$ and $\rho_i$ respectively -- i.e. the average number of bits needed to decode an object given a signal ($I_P$) and the averaged degeneracy of choices to name a given object ($I_{S}$).

			The low-vocabulary region B consists mostly of very polysemic signals (figure \ref{fig:02}{\bf c}). But codes with small vocabularies are not necessarily very polysemic -- e.g. along the Pareto front. Right next to region B, $I_P$ drops suddenly (area C in figure \ref{fig:02}{\bf c}) and then increases steadily as we tend towards the top right corner of $\Gamma$ (where a matrix sits with $a_{ij}=1 \> \forall i, j$). 

			Region B starts close to the star and it is also associated to a large synonymy index (figure \ref{fig:02}{\bf d}). This implies that $I_S$ increases sharply around the star as codes become less Pareto optimal. This swift increase does not happen if we start off anywhere else from the front. The condition for Pareto optimality is that codes do not have synonyms (see appendix \ref{app:01}), so this picture indicates that Pareto optimality degrades almost uniformly anywhere but near the star. This might have evolutionary implications: Languages around the B region require more contextual information to be disambiguated. That part of the morphospace might be difficult to reach or unstable if Pareto selective forces are at play.

		\subsection{Network structure}
			\label{sec:3.2}

			Words are not isolated entities within human language. Word inventories are only the first layer of language complexity. To make sense of language structure we need to consider how words interact, i.e. the patterns of connectivity associated to the underlying networks. Language networks can be defined in diverse ways \cite{Sole2010} by linking words together. It was early found that such networks are heterogeneous (the distribution of links displays very broad tails) and highly efficient in terms of navigation \cite{SoleSeoane2014}. The nature of these connections and the resulting graphs have been explored in very diverse classes of systems. Even the toy model studied here has been used to gain insight into the origins of complex linguistic features such as grammar and syntax \cite{FC2005, FC2006, Sole2005}. A network approach allows us to look at language from a system-level perspective, beyond the statistics associated to signal inventories. 

			\begin{figure}
				\begin{center}
					\includegraphics[width=\columnwidth]{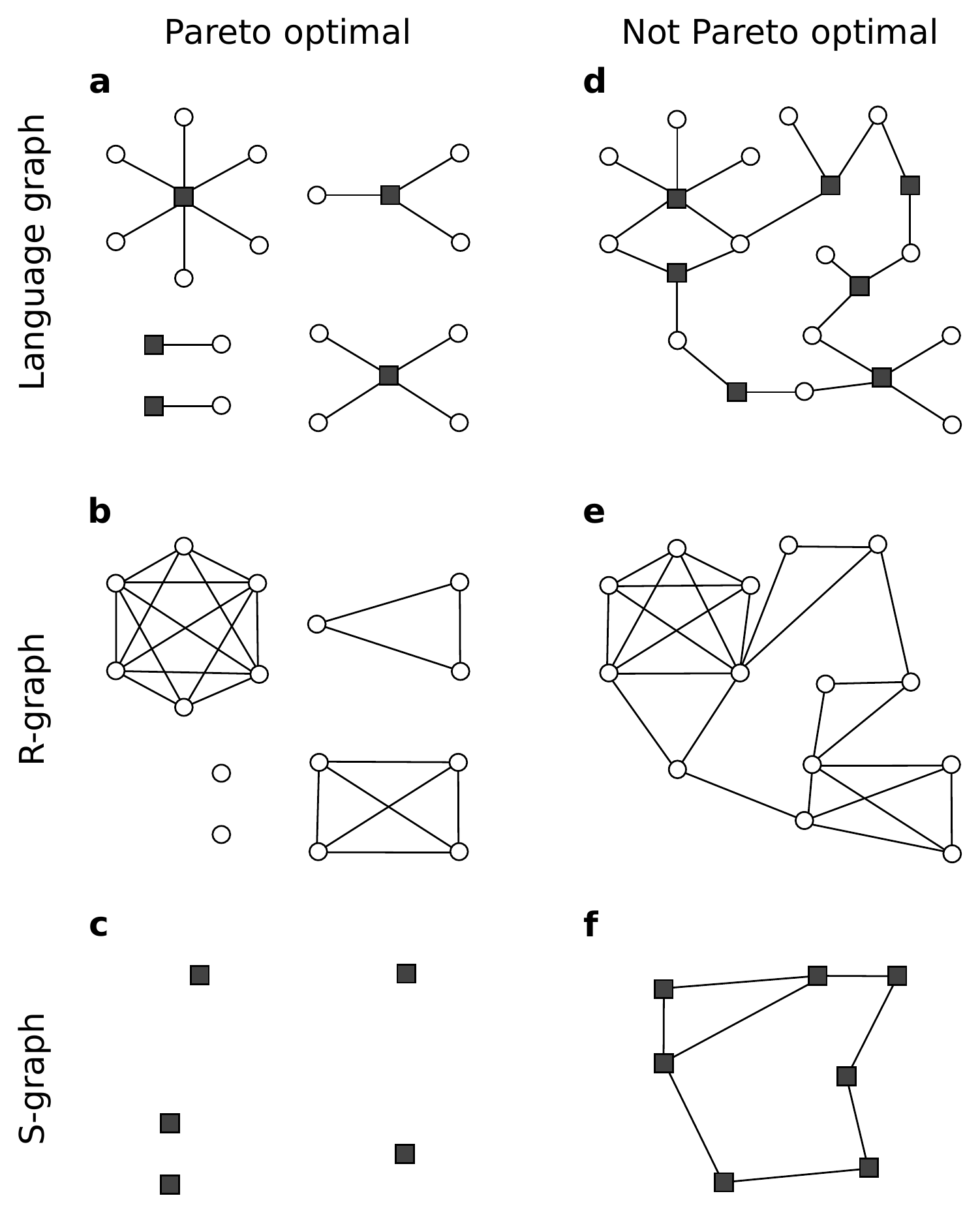}

          \caption{{\bf Different graphs derived from the language matrix.} {\bf a} A Pareto optimal language contains non-synonymous signals only. Its language graph consists of isolated clusters in which each signal clusters together a series of objects. {\bf b} Concepts within a cluster appear as cliques in the $R$-graph. {\bf c} The $S$-graph is just a collection of isolated nodes. {\bf d} Not Pareto optimal languages produce more interesting language graphs that might be connected (as here) or not. A connected language graph guarantees both a connected $R$- and $S$-graphs ({\bf e} and {\bf f} respectively).}

					\label{fig:03}
				\end{center}
			\end{figure}

			Each code in our model defines a bipartite network which connectivity is given by its matrix $A$ (figures \ref{fig:01}{\bf d-f} and \ref{fig:03}{\bf a} and {\bf d}). We refer to such a network as the {\em code graph}. We can derive two more networks from each code: one named {\em $R$-graph} (figure \ref{fig:03}{\bf b} and {\bf e}) in which objects $r_j, r_{j'} \in R$ are connected if they are associated to one same (polysemous) signal, and another one named {\em $S$-graph} (figure \ref{fig:03}{\bf c} and {\bf f}) in which signals $s_i, s_{i'} \in S$ are connected if they are synonymous. Because Pareto optimal codes do not contain synonyms, their bipartite code graphs consist of disconnected components in which the $i$-th signal binds together $\rho_i$ objects (figure \ref{fig:03}{\bf a}). Consequently, each Pareto optimal $R$-graph is a set of independent, fully connected cliques (figure \ref{fig:03}{\bf b}) and $S$-graphs are isolated nodes (figure \ref{fig:03}{\bf c}).

			\begin{figure*}
				\begin{center}
					\includegraphics[width=0.75\textwidth]{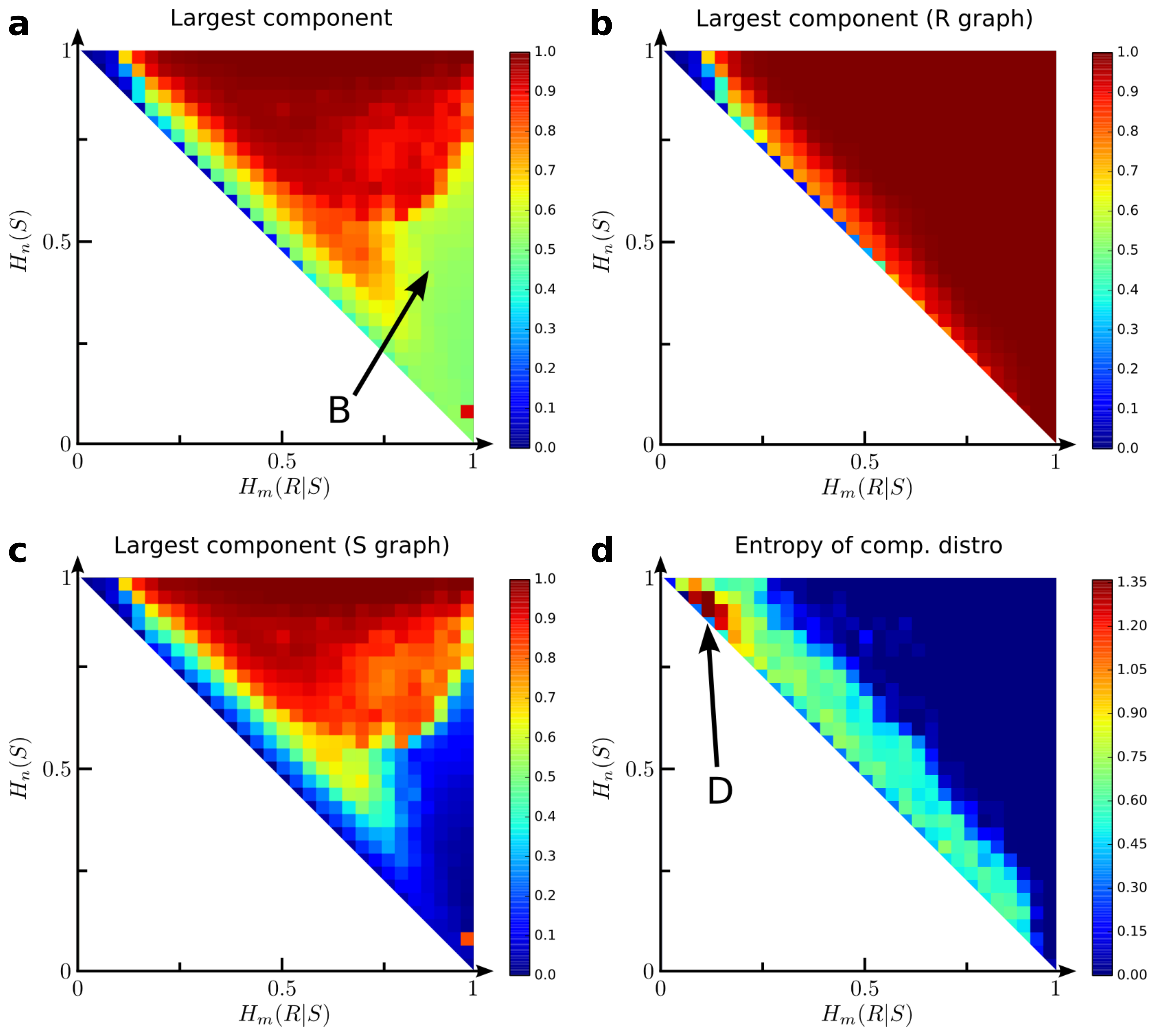}

          \caption{{\bf Network connectivity across the morphospace.} The size of the largest connected component is shown for code graphs ({\bf a}), $R$-graphs ({\bf b}), and $S-graphs$. {\bf d} Entropy of component size distribution is large around a band that runs parallel along the Pareto front. }

					\label{fig:04}
				\end{center}
			\end{figure*}

			A first characterization of network structure is the size of its most connected component. This is shown across the morphospace in figure \ref{fig:04}{\bf a-c} for code graphs, $R$-graphs, and $S$-graphs respectively. Regions with large connected components in their code graphs (figure \ref{fig:04}{\bf a}) largely overlap with networks with large effective vocabulary ($L$, figure \ref{fig:02}{\bf b}). The B region is the exception, as it displays an intermediate level of connectivity with very low $L$. This connectivity disappears for $S$-graphs in the B region, but the corresponding $R$-graphs remain very well connected. Hence a few signals are keeping together most of object space. Remarkably, $R$-graphs are very well connected everywhere throughout most of the morphospace, except for a very narrow region that extends from the one-to-one mapping along the Pareto front, more than halfway through it. 

			We kept track of the set of all connected components of a network $C \equiv \{C_i, i=1, \dots, N_C\}$ (with $N_C$ the number of independent connected components) and their sizes $||C_i||$. If $f(||C_i||)$ tells us the frequency with which components of a given size show up, then the entropy of this distribution 
				\begin{eqnarray}
					H_C &=& -{1\over ||N_C||} \sum_{i=1}^{N_C} f(||C_i||) log(f(||C_i||)) 
					\label{eq:08}
				\end{eqnarray}
			conveys information about how diverse the network is. This measure is shown in figure \ref{fig:04}{\bf d} for code graphs (it is virtually the same for $R$- and $S$-graphs). $H_C$ is small everywhere except on a broad band parallel to, and extending all along, the Pareto front. The fact that $H_C$ is so low in most of the morphospace stems from either one of three facts: i) Just one connected component exists, as in most of the area with large vocabulary. ii) Just a few signals make up the network, deeming all others irrelevant so that, effectively, all the features of the network can be summarized by a few archetypal graphs. iii) While a lot of signals are involved, they produce just a few different graphs. That shall be the case along the Pareto front (see appendix \ref{app:02.02}). 

			The band with moderate to large $H_C$ runs parallel to the Pareto front, but a little bit inside the morphospace. This would imply that, if the heterogeneity of the underlaying network were a trait selected for by human languages, they would be pulled off the Pareto front. Finally, $H_C$ is the highest around region D in figure \ref{fig:04}{\bf d}, at the end of the high-entropy band closer to the one-to-one mapping.

		\subsection{Complexity from codes as a semantic network}
			\label{sec:3.3}

			Words, concepts, and objects in the real world constitute an abstract semantic web whose structure shall be imprinted into (or stem from) our brains \cite{HuthGallant2012, HuthGallant2016}. It is often speculated that semantic networks must be easy to navigate. This in turn relates to the presence of a small-world underlying structure \cite{SteyversTenenbaum2005, SoleSeoane2014}. Navigation efficiency relates to system-level network properties. It would be interesting to quantify this using our codes as a generative toy model. 

			We approached this as follows. Starting with an arbitrary signal or object we implement a random walk moving into adjacent objects or signals. We record the nodes visited, hence generating symbolic strings associated to elements $r_j \in R$ and $s_i \in S$. The network structure shall condition the frequency $f(r_j)$ and $f(s_i)$ with which different objects and signals are visited. The entropies 
				\begin{eqnarray}
					H_R &=& -\sum_{j=1}^m f(r_j)log_m(f(r_j)), \nonumber \\
					H_S &=& -\sum_{i=1}^n f(s_i)log_L(f(s_i)). 
					\label{eq:10}
				\end{eqnarray}
			will be large if $R$ or $S$ are evenly sampled. They will present lower values if the network introduces non-trivial sampling biases. Hence, here low entropy is a measure of non-trivial structure arising form our toy generative model. We also recorded $2$-grams (couples of consecutive objects or signals during the random walk) and computed the corresponding entropies $H_{2R}$ and $H_{2S}$.

			This procedure is limited to sampling from the connected component to which the first node (chosen at random) belongs. If, by chance, we would land in a small connected component, these entropies would be artificially low disregarding of the structure that could exist elsewhere in the network. To avoid this situation we imposed that our generative model jumps randomly when an object was repeated twice since the last random jump, or since the start of the random walk. (We also interrupted the random walk when signals, instead of objects, were repeated. Results were largely the same.)

			\begin{figure*}
				\begin{center}
					\includegraphics[width=\textwidth]{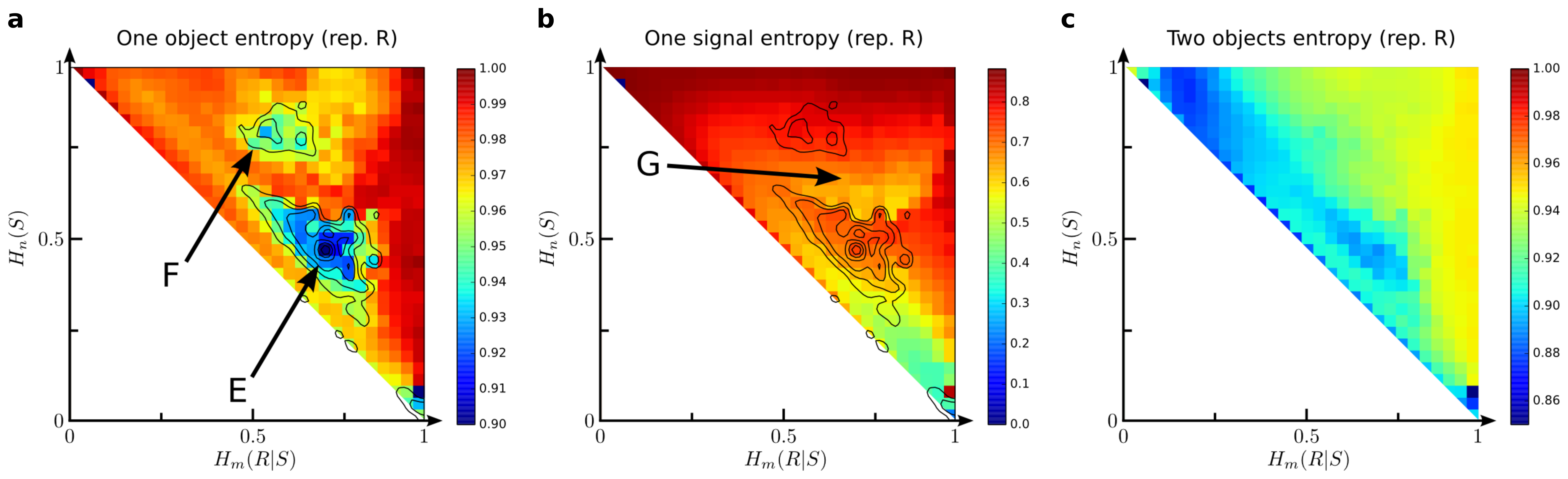}

          \caption{{\bf Complexity of codes as a random generative model.} {\bf a} Entropy of objects as sampled by a random walker ($H_R$) over the language network is close to its maximum throughout the morphospace, except for two non-trivial areas labeled E and F. Whichever mechanisms give rise to the heterogeneity there, they seem to be different, since the transition between E and F is not smooth. {\bf b} Entropy of signals as sampled by a random walker ($H_S$) is lower than its maximum across the morphospace, and the most singular areas do not correlate with the ones found for ($H_R$). Notably, region G seems to separate E and F and contain more heterogeneous signal sampling despite the largely homogeneous object sampling. {\bf c} $2$-grams of objects as sampled by a random walked present a lower entropy $H_{2R}$ than $H_R$, and only the E region seems to remain in place. }

					\label{fig:05}
				\end{center}
			\end{figure*}

			These measures present a non-trivial profile across the morphospace. We appreciate two regions in which $H_R$ drops (E and F in figure \ref{fig:05}{\bf a}). The code graphs around these areas must have some canalizing properties that break the symmetry between objects. However, the drop in entropy is of around a $10\%$ at most. (A third region with low $H_R$ near the star graph is discussed in appendix \ref{app:02.03} together with the measurements along the Pareto front.) 

			From figures \ref{fig:02}{\bf b} and \ref{fig:04}{\bf a}, region $E$ has moderately large vocabulary and size of connected component. It sits at a transition from lower values of these quantities (registered towards the front and within the B region) to the larger values found deeper inside the morphospace. Figure \ref{fig:04}{\bf d} shows how region E is located right out of the broad band with large $H_C$. All of this suggests that, within E, diverse networks of smaller size get connected into a large component which inherits part of the heterogeneous structure. This heterogeneity results in a bias in the sampling of objects, but not as much in the sampling of signals. The lowest $H_S$ is registered towards the start-graph instead (see appendix \ref{app:02.03}). Note also that biases in signal sampling are larger (meaning lower $H_S$) throughout the morphospace -- compare the scale of the color bars in figures \ref{fig:05}{\bf a} and {\bf b}.

			Region F sits deeper inside the morphospace, where vocabulary size is almost the largest possible and the connected component involves most of all signals and objects. The network here is well consolidated suggesting that the bias of object sampling comes from non-trivial topologies established through redundant paths. Interestingly, regions E and F are separated by an area (G in figure \ref{fig:05}{\bf b}) with a more homogeneous sampling of objects and a relatively heterogeneous sampling of signals. $H_S$ within F itself is larger than in G, suggesting no remarkable bias on word sampling in F despite the bias on object sampling, and vice-versa. We take all this as an example of the diversity found in the morphospace, which allows an important asymmetry between words and objects inducing heterogeneity in one set while keeping the other homogeneous. 

			Figure \ref{fig:05}{\bf c} shows $H_{2R}$, the entropy of $2$-grams objects produced by the sampling. It seems to inherit a faded version of the E region from $H_R$. It is also low along a band largely overlapping the one shown in figure \ref{fig:04}{\bf d} for $H_C$. The largest drop in $H_{2R}$ happens closer to the one-to-one mapping. It makes intuitive sense that codes in this last area start consisting of networks similar to the one-to-one mapping in which extra words connect formerly isolated objects, hence resulting in a bias of couples of objects that appear together. The entropy of $2$-gram words ($H_{2S}$, not shown) is largely similar to that of $H_S$ (figure \ref{fig:05}{\bf b}).

		\subsection{Zipf, and other power laws} 
			\label{sec:3.4}

      Zipf's law is one of the most notable statistical patterns in human language \cite{Zipf1949}. Despite important efforts \cite{CorominasSole2010, CorominasSole2011, CorominasMurtraSole2016}, the reasons why natural language should converge towards this distribution of word frequencies are far from definitive. Detailed research of diverse written corpora suggests that under certain circumstances (e.g. learning children, military jargon, cognitively impaired patients) the frequency of words presents a power-law distribution with a generalized exponent \cite{Ferrer2005, BaixeriesFerrer2013}. 

      In the past, different authors have studied how well the least-effort toy model can account for Zipf's distribution of words \cite{FerrerSole2003, ProkopenkoPolani2010, SalgeProkopenko2013}. Assuming that every object needs to be recalled equally often, and that whenever an object $r_j$ is recalled we choose uniformly among all the synonymous words naming $r_j$; we can compute the frequency with which a word would show up given a matrix $A$. This is far from realistic: not all objects need to be recalled equally often, and not all names for an object are used indistinctly. This does not prevent numerical speculation about computational aspects of the model, which might also be informative about the richness of the morphospace. 

      The first explorations of the model \cite{FerrerSole2003} indicated that Zipf's law lays just at the transition point between the star and one-to-one codes. This suggested that self-organization of human language at the least-effort critical point could be a driving force for the emergence of Zipf's distribution in word corpora. Later on, it was shown analytically that while it is possible to find languages owing Zipf's law at that transition, this is not the most frequent distribution among Pareto optimal languages \cite{ProkopenkoPolani2010, SalgeProkopenko2013}. This is consistent with the diversity that we find at the critical manifold (see appendix \ref{app:02}). This also implies that if Pareto-optimal least-effort is a driving force of language evolution, it would not be enough to constrain the word distribution to be Zipfian. Other authors \cite{FortunyCorominas2013} have provided mathematical arguments to expect that Zipf's law will be found right at the center of the Pareto front (with $\Omega_h = 1/2 = \Omega_s$). Again, even if human language would converge to this singular point, this shall still leave the word distribution unconstrained. 

			\begin{figure*}
				\begin{center}
					\includegraphics[width=\textwidth]{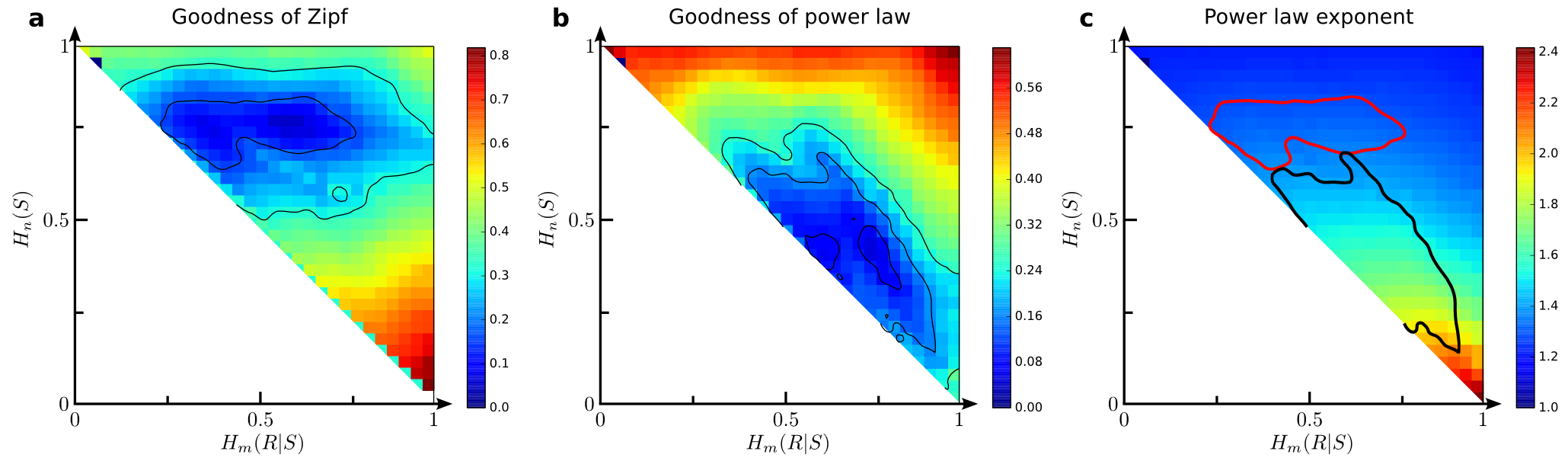}

          \caption{{\bf Power laws from the least-effort model.} {\bf a} Goodness of fit of the word distribution from the toy, least-effort model to a Zipf law. {\bf b} Goodness of fit of the word distribution from the model to an arbitrary power law. {\bf c} Exponent obtained when fitting the word distribution of the model to the arbitrary power law from panel {\bf b}. In each case, the level curves indicate areas where a Kolmogorov-Smirnov test suggest a good fit. }

					\label{fig:06}
				\end{center}
			\end{figure*}

      We built the word frequencies from each $A$ matrix using the prescriptions just outlined (all objects are referred to equally often, all synonyms are used indistinctly). To asses how well each distribution is explained by Zipf's law, we used a Kolmogorov-Smirnov (KS) test (scores are plotted in figure \ref{fig:06}{\bf a}). The area with better fitness to Zipf is broad and stretches notably inside the morphospace, indicating that Zipf's distribution does not necessarily correlate with least-effort. This area runs horizontally with values $H_n(S) \sim 0.75$ and roughly $H_m(R|S) \in (0.25, 0.75)$. In the best (least-effort) of cases, speakers incur in costs ($\Omega_s \equiv H_n(S)$) three times higher that hearers. Less Pareto optimal codes that achieve Zipf always have a greater cost associated to speakers too. 

      Following the methods in \cite{ClausetNewman2009}, we fitted the word frequencies of each $A$ matrix to power laws with arbitrary exponents. The KS-score from figure \ref{fig:06}{\bf b} reveals an alternative region with large goodness of fit that runs parallel along the lower part of the Pareto front. However, the exponent obtained through this method (figure \ref{fig:06}{\bf c}) falls around the $1.6-1.8$ region, far from Zipf's law. Our morphospace seems a powerful tool to plot the diverse exponents found in special written corpora \cite{Ferrer2005, BaixeriesFerrer2013}. This could provide insights about how the language network structure changes in those cases. 

      These numerical findings present notable evidence against least-effort as an explanation of Zipf's law. Not Pareto-optimal codes exist with larger fitness to Zipf's than least-effort languages (figure \ref{fig:06}{\bf a}) and codes along the critical manifold seem better fitted by other power laws (see appendix \ref{app:02.04}, figure \ref{fig:11}{\bf b} and {\bf c}). Two important limitations of the model should be considered: First, objects and synonyms are not equally frequently used. Introducing asymmetries (hopefully realistic ones, derived from actual word usage) could alter the balance between hearer and speaker efforts. Second, we are dealing with relatively small matrices ($200 \times 200$) to make the computations tractable. Good measurements of power-law exponents demand larger matrices. Alleviating these handicaps of the model shall bring back evidence supporting the least-effort principle.

	\section{Code archetypes and real languages}
		\label{sec:4}

		\begin{figure*}
			\begin{center}
				\includegraphics[width = \textwidth]{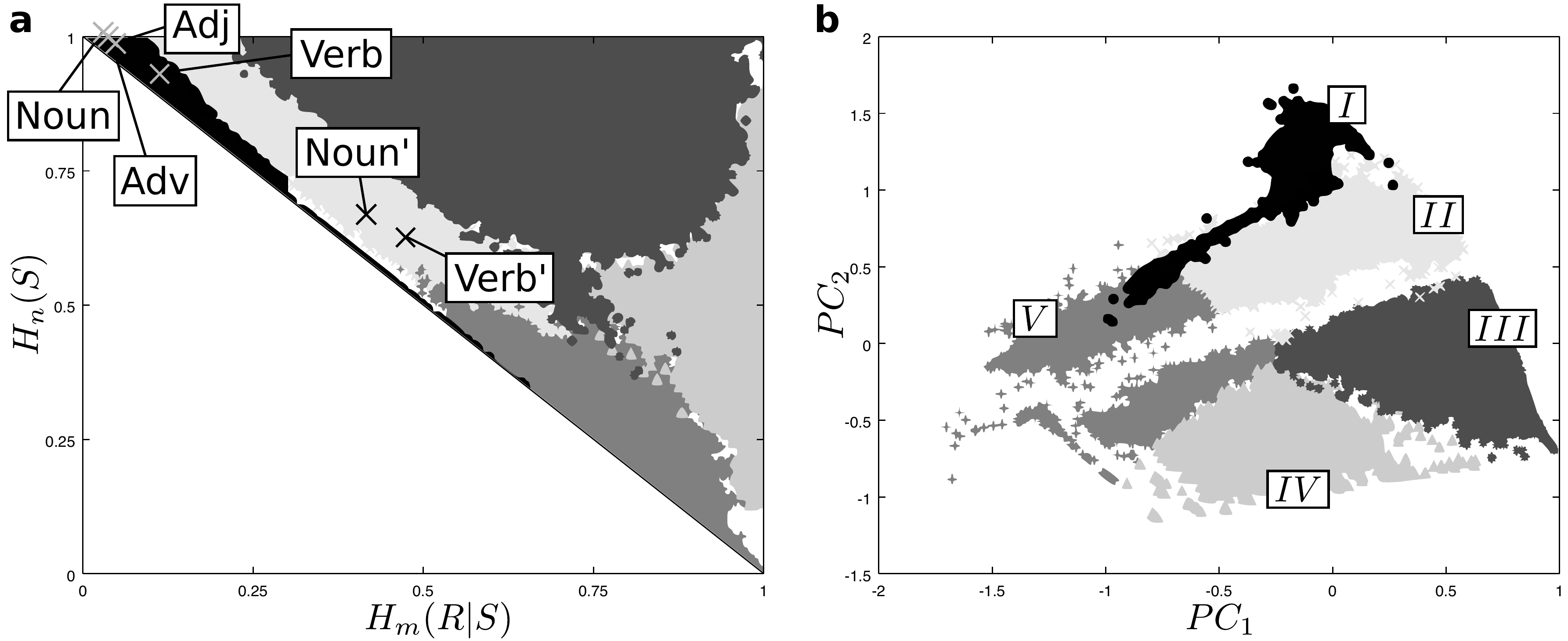}

        \caption{{\bf Clustering of languages across the morphospace. } $k$-means clustering using all principal components reveals a consistent structure in the morphospace. Five cluster are shown here. Real languages fall within cluster $I$, close to the one-to-one mapping proper of animal communication systems. The real matrices are marked: {\em Adj} for the adjectives, {\em Adv} for the adverbs, {\em Noun} for the nouns, and {\em Verb} for the verb. If certain grammatical words are included (named with an apostrophe: {\em Noun'} for nouns and {\em Verb'} for verbs) they move into cluster $II$ and towards the center of the morphospace, relatively close to the Pareto front. {\bf b} All clusters get further segregated in two principal component space. This space appears interrupted by a stripe along which no codes exist. }
				\label{fig:07}
			\end{center}
		\end{figure*}

		We introduced different measurements over the matrices $A$ of our toy model. The emerging picture, far from a smooth landscape, is that the language morphospace breaks into finite, non-trivial ``archetypes''. To support this, we ran additional analyses to discern relevant dimensions for our problem. With all the measurements described above we moved into Principal Component (PC) space. $5$ PCs we needed to explain $90\%$ of the variation in the data. We then applied a $k$-means algorithm \cite{Lloyd1982} using all PC values. For $k=5$, running the algorithm several times we converged consistently upon similar clusters that we classify as follows (figure \ref{fig:07}, clockwise from top-left): 
			\begin{itemize}

        \item[$I$] Codes near the one-to-one mapping and upper two thirds of the Pareto front. This includes the graphs with largest $H_C$ (figure \ref{fig:04}{\bf d}).

        \item[$II$] Codes along a stripe parallel to the upper half of the Pareto front. This overlaps largely with the region with large $H_C$ (figure \ref{fig:04}{\bf d}) and low $H_{2R}$ (figure \ref{fig:05}{\bf c}). 

        \item[$III$] Bulk interior region consisting mostly of codes with a single connected component and large vocabulary. It includes region F with low $H_R$ (figure \ref{fig:05}{\bf a}). 
			
        \item[$IV$] Region B from figure \ref{fig:02}{\bf b-d}, consisting of codes with large polysemy and small vocabularies. These demand exhaustive contextual cues for communication.

        \item[$V$] Codes along the lower half of the Pareto front and a thick stripe parallel to it. This overlaps partly with the region with good fit to power-laws (figure \ref{fig:06}{\bf b}). \\

			\end{itemize}

    Solutions to the original least-effort problem were widely analyzed in the literature from a theoretical perspective. These studies focused on the model's phase transition \cite{FerrerSole2003}, on the existence of Zipf's distribution at the critical point \cite{FerrerSole2003, ProkopenkoPolani2010, SalgeProkopenko2013, SoleSeoane2014}, or on mechanisms that could drive languages to this distribution \cite{Ferrer2005, FortunyCorominas2013, SeoaneSole2015b}. Based on such analyses it was speculated that human language should lay at the transition point, since either extreme was not suitable to describe the flexibility of our communication systems. One-to-one mapping, associated to animal codes, was deemed rather rigid and memory demanding. This raised a point that ambiguity would be the price to pay for least-effort efficient language. On the other hand, the star code makes communication impossible unless all the information is contextually explicit. 

    The assessment of real languages using this toy model is missing in the literature. This owes, perhaps, to the difficulty of building matrices $A$ out of linguistic corpora. WordNet \cite{Fellbaum1998, Miller1995} contains a huge database with different semantic relationships, including manually annotated relationships between words and objects or concepts. A few examples:
			\begin{lstlisting}
ape (...) 02470325 09964411 09796185	
car (...) 02958343 02959942 02960501 ...
complexity (...) 04766275	
rugby (...) 00470966	
			\end{lstlisting}
		The parentheses stand for additional information not relevant here. Each word is associated to several codes. Each code identifies a unique, unambiguous object or concept. For example, $02959942$ refers to the car of a railway while $02960501$ refers to the gondola of a funicular. The word ``car'' appears associated to these two meanings among others. WordNet makes this information available for four separate grammatical categories: adjectives, adverbs, nouns, and names.

    We built the corresponding $A$ matrices out of this database and evaluated $H_m(R|S)$ and $H_n(S)$ for each grammatical category. All four categories contain more signals than objects, hence synonyms exist and languages are not Pareto optimal. Theoretical models (also beside ours) argue that synonyms should not exist in optimal codes \cite{NowakKrakauer1999, FerrerSole2003, SalgeProkopenko2013}, but they seem real in folk language. Synonymy shall also have degrees, with linguists dissenting about whether two terms name the precise same concept. Such information is lost due to our coarse mapping into binary matrices, but it is possible to extend our analysis if $A$ displayed likelihoods $a_{ij} \in [0, 1]$ indicating affinity between words and concepts. 

    Figure \ref{fig:07}{\bf a} shows all grammatical categories (labeled {\em Adj}, {\em Adv}, {\em Noun}, and {\em Verb} respectively) in our morphospace. While not Pareto optimal, they appear fairly close to the front. They also appear near the one-to-one mapping. This would suggest that human language is not such a great departure from codes associated to other animals, thus contradicting several arguments in least-effort literature. Also, all matrices appear restricted to a small area, leaving the huge morphospace mostly unexplored.

    However, the WordNet database does not contain grammatical words such as pronouns. Some proper names appear in the {\em Noun} database (e.g. Ada and Darwin), but `she', `he', or `it' are not included. Any feminine proper name can be substituted by `she', while `it' can represent any common noun. Similarly, in English most verbs can be substituted by `to do' or `to be' -- e.g. ``She plays rugby!'' becomes ``Does she play rugby?'' and eventually ``She does!''. Appending these words to the corresponding matrices would account for adding signals that can name almost every object. We simulated this by adding a single word to the real matrices for nouns and verbs that can name any other concept. This changed the corresponding $H_m(R|S)$ and $H_n(S)$ values, shifting these codes right into the central-lower part of cluster $II$ (figure \ref{fig:07}{\bf a}, points marked {\em Noun'} and {\em Verb'} with apostrophes), near the center of the Pareto front. This suggests that grammatical words might bear all the weight in opening up the morphospace for human languages, with most semantic words conforming a not-so-outstanding network close to the one-to-one mapping and still demanding huge memory usage.

	\section{Discussion}
		\label{sec:5}

		The least-effort model discussed in this paper has long captured the attention of the community. It features a core element of most communication studies -- namely, the ``coder to noisy-channel to decoder'' structure found in Shannon's original paper on information theory \cite{Shannon1948}, as well as in more recent experiments on the evolution of languages \cite{Kirby2001, KirbySmith2008, Steels2015}. This toy model allows us to formulate a series of questions regarding the optimality of human language and other communication systems. These had been partly addressed numerically \cite{FerrerSole2003} and analytically \cite{ProkopenkoPolani2010, SalgeProkopenko2013}. It was found that a first order phase transition separates the one-to-one mapping from a fully degenerated code. It was further speculated that a critical point existed at this transition, and that human language may be better described by that regime owing to the repertoire of properties of critical systems \cite{FerrerSole2003}. However, this hypothesis has never been confronted with empirical data. The criticality at the phase transition was never settled either. Finally, by looking only at least-effort languages the vast majority of codes present in the model was left unexplored. 

		This paper uses a formalism grounded on Pareto optimality to recover the first order phase transition of the model \cite{SeoaneSole2013, SeoaneSole2015b, Seoane2016} and to prove analytically that it indeed contains a critical point \cite{SeoaneSole2015c}. Besides, the paper characterizes the very rich morphospace of communication codes beyond the optimality constraints. Finally, it addresses for the first time empirically the hypothesis about the optimality and criticality of human language within the least-effort model. 

		The language morphospace turns out to be surprisingly rich, far from a monotonous variation of language features. Different quantities such as the synonymy of a code, its network structure, or its ability to serve as a good model for human language (e.g. by owing Zipf's law) present non-trivial variations across the morphospace. These quantities might or might not align with each other or with gradients towards Pareto optimality, and may hence pose newer conflicting forces that human language or other communication systems shall be driven by. 

		To portray real human languages within the least-effort formalism we resorted to the WordNet database \cite{Miller1995, Fellbaum1998}. Raw matrices extracted from this curated directory locate human language close enough to one-to-one mappings proper of other animals, and in the interior of the morphospace. This would invalidate the previous hypothesis that human language belongs far apart from animal communication and along the critical point of the model. But introducing grammatical particles such as the pronoun `it' or the auxiliary form of the verb `to do' (both missing from the WordNet database) does move human language far away from one-to-one mappings and closer to the center of the critical manifold. Both found locations for human languages (before and after adding grammatical particles) present some interesting properties such as a large entropy of concept-cluster size ($H_C$, figure \ref{fig:04}{\bf d}). This quantity drops to zero at the Pareto front, suggesting evolutionary forces that could pull real languages away from the kind of least-effort optimality studied here. 

		Our results suggest a picture of human language consisting of a few referential particles operating upon a vastly larger substrate of otherwise unremarkable words. The transformative power of grammatical words is further highlighted if we consider that just one was enough to completely displace human codes into a more interesting region of the morphospace. This invites us to try more refined versions of the model in which grammatical particles are introduced with more care -- e.g. based on how often pronouns substitute another word daily language usage. This also poses interesting questions regarding the sufficient role of such grammatical units to trigger and sustain full-fledged language. 

		The WordNet database is only the most straightforward possibility to map human language into the model. Controlled experiments or recent neuroscientific developments \cite{HuthGallant2016} offer new opportunities to validate or challenge our results or to address new questions in evolutionary or developmental linguistics. In this sense, the morphospace introduced here offers an elegant framework upon which to trace the progression, e.g., of synthetic languages grown in the lab \cite{Kirby2001, KirbySmith2008} or in silico \cite{Steels2015}; or to depict other signal-object mappings found in culture or biology, such as the `codon-amino acid' correspondence of the genetic code.

	\vspace{0.2 cm}

	\section*{Acknowledgments}
	
		The authors thank the members of the Complex Systems Lab for useful discussions. This study was supported by the Botin Foundation, by Banco Santander through its Santander Universities Global Division, the support of Secretaria d'Universitats i Recerca del Departament d'Economia i Coneixement de la Generalitat de Catalunya and by the Santa Fe Institute.

	\vspace{0.2 cm}

	\section*{Author contributions}
	
		Both authors developed the ideas presented here, designed the computational experiments, and composed the text. LFS performed the simulations and data analyses. The cursed ghost of Mary Shelley hunted us till today.

\appendix 
	
	\section{Phase transitions, criticality, and sampling of the model's design space} 
		\label{app:01}

		Given a number $n$ of signals and a number $m$ of objects, the set of all $n \times m$ binary matrices constitutes the {\em design space} $\Gamma$ of our toy model. Each matrix $A$ has a pair of costs $\left(\Omega_h(A), \Omega_s(A)\right) \equiv \left( H_m(R|S), H_n(S) \right)$ that map $\Gamma$ into the $2$-D plane. These costs are optimization targets of the MOO least-effort problem, so we often refer to the $\Omega_h-\Omega_s$ place as {\em target space}. Here we set up to explore the overall shape of our design space in target space, and what consequences this has for the model from an optimality viewpoint. 

		A first step is to find the extent of $\Gamma$ in the $\Omega_h-\Omega_s$ plane. The global minima of $\Omega_h$ and $\Omega_s$ delimit two of the boundaries of $\Gamma$. Take the matrix associated to the minimal hearer effort, $A_h \equiv I_n$, where $I_n$ denotes the $n \times n$ identity matrix so that $a_{ij} = \delta_{ij}$ (with $\delta_{ij}=1$ for $i=j$ and zero otherwise, figure \ref{fig:01}{\bf c}). This matrix minimizes the effort for a hearer: signals are not degenerated and she does not need to struggle with ambiguity. Naturally, $\Omega_h(A_h) = 0$ while from equation \ref{eq:04} $\Omega_s(A_h) = log_{n}(m)$. So $A_h$ dwells on the top-left corner of the set of possible languages in target space. Consider on the other hand $A = A_s \equiv \{a_{ij} = \delta_{ik}\}$, where $k$ is an arbitrary index $k \in [1, n]$. Here one given signal ($s_k$) is used to name all existing $r_j$ resulting in the minimal cost for the speaker. It follows from equations \ref{eq:03} and \ref{eq:04} that $\Omega_h(A_s) = 1$ and $\Omega_s(A_s) = 0$, so this matrix sits on the bottom-right corner of $\Gamma$. Owing to the graph representing $A_s$ (figure \ref{fig:01}{\bf d}) we refer to it as the star graph. 

		These optimal languages for one of the agents also suppose the worst case for its counterpart. Hence, (for $n = m$) no matrices lay above $\Omega_s = log_{n}(m)$ nor to the right of $\Omega_h = 1$. A language with as many signals as objects and with all of its signals completely degenerated sits on the upper right corner of the corresponding space. This is encoded by a block matrix filled with ones. For simplicity, the vertical axis in all figures of this paper has been rescaled by $log_m(n)$ so that the horizontal boundary of the set is $\Omega_s = 1$. (This happens naturally if $n=m$, which we take often to be the case.)

		The only boundary left to ascertain is the one connecting $A_h$ and $A_s$ in the lower left region of target space. This constitutes the optimal tradeoff when trying to simultaneously minimize both $\Omega_h$ and $\Omega_s$, hence it is the Pareto front ($\Pi_{\Gamma}$) of the multiobjective least effort language problem. It can have any shape as long as it is monotonously decreasing (notably, it does not need to be derivable nor continuous), and its shape is associated to phase transitions and critical points of the model \cite{SeoaneSole2013, SeoaneSole2015a, SeoaneSole2015b, SeoaneSole2015c, Seoane2016}. 

		Prokopenko et al. \cite{ProkopenkoPolani2010, SalgeProkopenko2013} computed analytically the global minimizers of equation \ref{eq:02}. These turn out to be all matrices $A$ that do not contain synonyms -- i.e. which have just one $1$ in each column. For those codes, using some algebra we come to the next expressions for the target functions:
			\begin{eqnarray}
				\Omega_h &\equiv& H_m(R|S) = \log_{m}(n) \sum_{i=1}^n {\rho_i \over m} log_{n} (\rho_i), \\
				\label{eq:05.1}
				\Omega_s &\equiv& H_n(S) = \log_{n}(m) - \sum_{i=1}^n {\rho_i \over m} log_{n} (\rho_i), \\ 
				\label{eq:05.2}
				\Omega_s &=& \log_{n}(m) - {1 \over log_{m}(n)} \Omega_h; 
				\label{eq:05.3}
			\end{eqnarray}
		where $\rho_i$ is the number of objects named by the $i$-th signal. Equation \ref{eq:05.3} defines a straight line in target space (figure \ref{fig:02}{\bf a}). It can be shown that minimizers of equation \ref{eq:02} are always Pareto optimal \cite{SeoaneSole2013, Seoane2016}. The opposite is not necessarily true (there might be Pareto optimal solutions that do not minimize equation \ref{eq:02}), but the curve from equation \ref{eq:05.3} connects $A_h$ and $A_s$ in target space exhausting any other possibility. In this problem there cannot exist other Pareto optimal matrices and equation \ref{eq:05.3} constitutes the whole MOO solution by itself. 

		Assuming $n=m$, $\Pi_{\Gamma}$ is the straight line $\Omega_s = 1 - \Omega_h$ (figure \ref{fig:02}{\bf a}). This implies that the global optimizers of equation \ref{eq:02} undergo a first order phase transition at $\lambda = \lambda_c \equiv 1/2$ \cite{SeoaneSole2013, SeoaneSole2015b, Seoane2016}, thus confirming previous observations about the model \cite{FerrerSole2003, ProkopenkoPolani2010, SalgeProkopenko2013}. In the literature it is also speculated that this phase transition has a critical point, but this could not be confirmed. This is precisely what is predicted for MOO problems whose Pareto front is a straight line, so equation \ref{eq:05.3} proves the critical nature of the system analytically. Besides, a straight Pareto front implies that any Pareto selective force\footnote{Defined by \cite{SeoaneSole2015c} as any algorithmic procedure that drives the system towards its Pareto front.} will poise the system to its critical state \cite{SeoaneSole2015c}. \\

		Again assuming $n=m$, the triangle shown in figure \ref{fig:02}{\bf a} contains all possible communication codes according to our model. For a modest $n = 200$  there are $2^{nm} = 2^{40000}$ possible codes. In section \ref{sec:3} we report a series of measurements taken on language networks throughout the morphospace. For these to be representative we need that $\Gamma$ is sampled evenly across the $\Omega_h-\Omega_s$ plane. Several strategies were tried with that aim, such as wiring objects to signals with a low probability $p$, generating a few Pareto optimal codes, the star and the one-to-one mappings, mutations and combinations of these, etc. This approach allowed to sample very small and isolated regions of the morphospace. To improve over this, we implemented a genetic algorithm with $N_s = 10\>000$ matrices that would proceed until the upper-right half of a $30 \times 30$ grid in $(\Omega_h, \Omega_s) \in [0, 1]\times[0, 1]$ was evenly covered with roughly $20$ matrices in each square of the grid. Going beyond $n = 200 = m$ proved to be computationally very costly. 

		This cost could be partly alleviated for Pareto optimal matrices. These are defined as languages that do not contain synonyms. This allowed a sparse encoding of these matrices. Some computations were also simplified (e.g. the costs are bound by equation \ref{eq:05.3}). Because of this, we could perform an alternative sampling of $N_s = 10\>000$ matrices along the Pareto front with more signals and objects (up to $1\>000$). Different stochastic mechanisms were used to seed a similar genetic algorithm that ensured an even sample of matrices along the front. While Pareto optimal matrices always included $1\>000$ objects, some of the mechanisms to generate them would result in languages with less signals. In the following, all quantities have been properly normalized for comparison. The results of the different measurements on Pareto optimal matrices are reported in appendix \ref{app:02}. 

		The fact that simple recipes to build matrices (and mutations thereof) resulted in a poor sampling of our language morphospace provides some relevant insight about how difficult it is to access most of $\Gamma$. In order to sample the whole space we needed non-trivial algorithms and a target that the whole space was covered. If we would observe actual languages in singular regions of the morphospace, we could wonder about what evolutionary forces brought those languages there and suggest that more is needed than what simple rules offer for free.

	\section{Complexity of language networks along the Pareto front} 
		\label{app:02}

		In section \ref{sec:3} we reported a series of measurements taken over an even sample across the morphospace. Those results are complemented here by measurements taken over a more exhaustive sample of the Pareto front which includes larger matrices (with up to $1\>000$ signals and objects, as opposed to the $n=400=m$ in the main text). In the following sections we analyze the same measurements of vocabulary, network structure, matrix as a generative model, and goodness of fit to power-law that we analyzed above. 

		The critical manifold is just a straight line, which allows us to present simpler plots. Below, the horizontal axis reports the value of $\Omega_h \equiv H(R|S)$ along the front. This is, the one-to-one mapping lays at the leftmost part of the plot and the star graph at the rightmost end. 

		\subsection{Characterizing the vocabulary}
			\label{app:02.01} 

			\begin{figure}
				\begin{center}
					\includegraphics[width = \columnwidth]{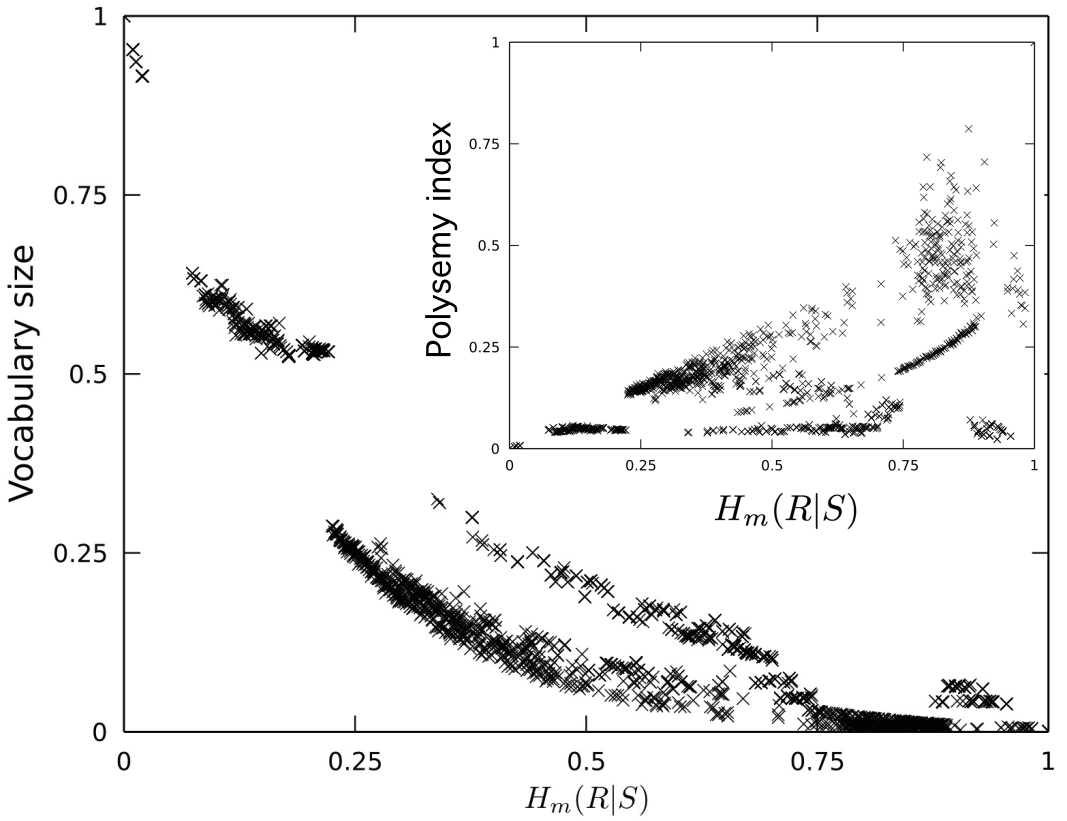}

          \caption{{\bf Vocabulary size and polysemy along the Pareto front.} {\bf a} Codes along the Pareto front keep a relatively low vocabulary except close to the one-to-one mapping. Also, two branches seem noticeable around the middle of the front, suggesting that similar Pareto optimal values of $H_m(R|S)$ and of $H_n(S)$ can be achieved with differently wired codes. {\bf b} A reduced vocabulary size does not result in a strictly monotonous increase of polysemy as we approach the star code. Instead, languages with similar $H_m(R|S)$ may present different polysemy levels. The range available grows as we approach the maximally ambiguous code. }
					\label{fig:08}
				\end{center}
			\end{figure}

			By definition, Pareto optimal languages have no synonyms hence $I_S = 0$. We report next vocabulary size ($L$) and polysemy index ($I_P$) along the front. 

			Figure \ref{fig:08} shows that the effective vocabulary size does not decrease linearly as we proceed from the one-to-one mapping ($L=n$) to the star ($L=1$). Furthermore, at most given points along the front, there seem to be several languages with the same effort for both speaker and hearers, and yet with different vocabulary size. This indicates that there are different strategies to achieve the same degree of optimality, or that being Pareto optimal leaves the diversity of languages largely unconstrained. 

      Regarding polysemy, we could also expect that it would build up uniformly as we approach the star code. Instead we see that at each point along the front there are very different codes showing a range of polysemy (figure \ref{fig:08}, inset). The maximum of this range does grow with $H_m(R|S)$, but we know that $I_P$ has to be maximum and unique for the star graph. The fact that similar Pareto optimal codes present such diverse $I_P$ (as well as $L$) suggests a great diversity within the critical point of the model. We will find that this is a recurrent theme of Pareto optimal languages for other measurements as well. 

	  \subsection{Network structure}
	  	\label{app:02.02}

	  	We recall now the bipartite network structure (code graph) and the corresponding $R$- and $S$-graphs in object and signal space. These are naturally induced by the $A$ matrices as illustrated in figure \ref{fig:03}. Associated to them, we report the size of the largest connected component ($||C_1||$) for each graph, and the entropy of the distribution of component sizes ($H_C$) as introduced in section \ref{sec:3.2}. 

			\begin{figure*}
				\begin{center}
					\includegraphics[width=0.75\textwidth]{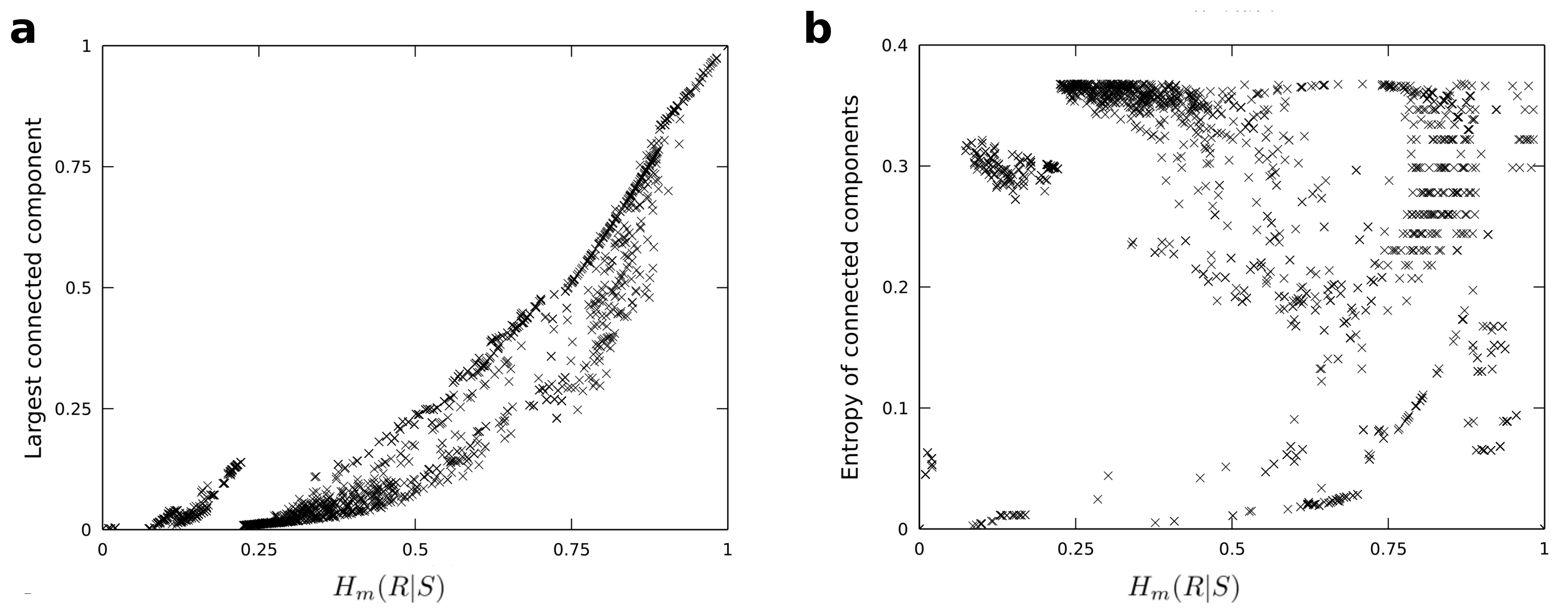}

          \caption{{\bf Network connectivity along the Pareto front.} {\bf a} Along the front, the size of the largest connected component grows from $1/m$ to $1$ as we move from the one-to-one mapping to the start graph. {\bf b} The entropy of component size distribution shows a large degree of degeneracy even for single points along the front. }

					\label{fig:09}
				\end{center}
			\end{figure*}

			For languages along the Pareto front, the largest connected component of the $S$-graph has trivially just $1$ signal (because, again, there are no synonyms). This implies that the largest connected component of the code and $R$-graphs are virtually the same. Figure \ref{fig:09}{\bf a} shows the size (normalized to the maximum value possible) of the largest component of the code graph along the front. It grows as we move from the one-to-one mapping to the star code, but this growth is again mostly non-linear and often several possibilities coexist at each point of the front. 

			Regarding $H_C$, we find no consistent pattern throughout the front (figure \ref{fig:09}{\bf b}). This is the measure for which we find less correlation along any direction in Pareto optimal codes, again suggesting that the diversity of networks along the front is largely unconstrained. Notwithstanding, this variability is perhaps not so salient: $H_C$ here is small as in most of the morphospace (compare the scale in the color bar of panel \ref{fig:04}{\bf d} against the vertical axis of panel \ref{fig:09}{\bf b}). Moving apart from the star graph, we know that several signals are involved in Pareto optimal languages (as the vocabulary size implies -- figure \ref{fig:08}) and yet $H_C$ is kept low and relatively constant throughout. This suggests that, while a lot of disconnected components coexist to make up a Pareto optimal language network, their sizes are similar resulting in just a few graphs similar to each other.

		\subsection{Complexity from codes as a semantic network}
			\label{app:02.03}

			We turn our attention now to language matrices as generative toy models of semantic relationships. Therefore, we had introduced a random walk over code graphs in section \ref{sec:3.3}. These allowed us to capture, with a series of entropies ($H_{R,S}$ and $H_{2R,2S}$), whether the network structure somehow biased the sampling of signals or objects as it traversed the network randomly. Large entropies in the distribution of sampled objects or signals implied networks that do not induce remarkable structures. Meanwhile, noteworthy biases in object or signal sampling would result in lower entropies than expected. 

			\begin{figure*}
				\begin{center}
					\includegraphics[width=0.75\textwidth]{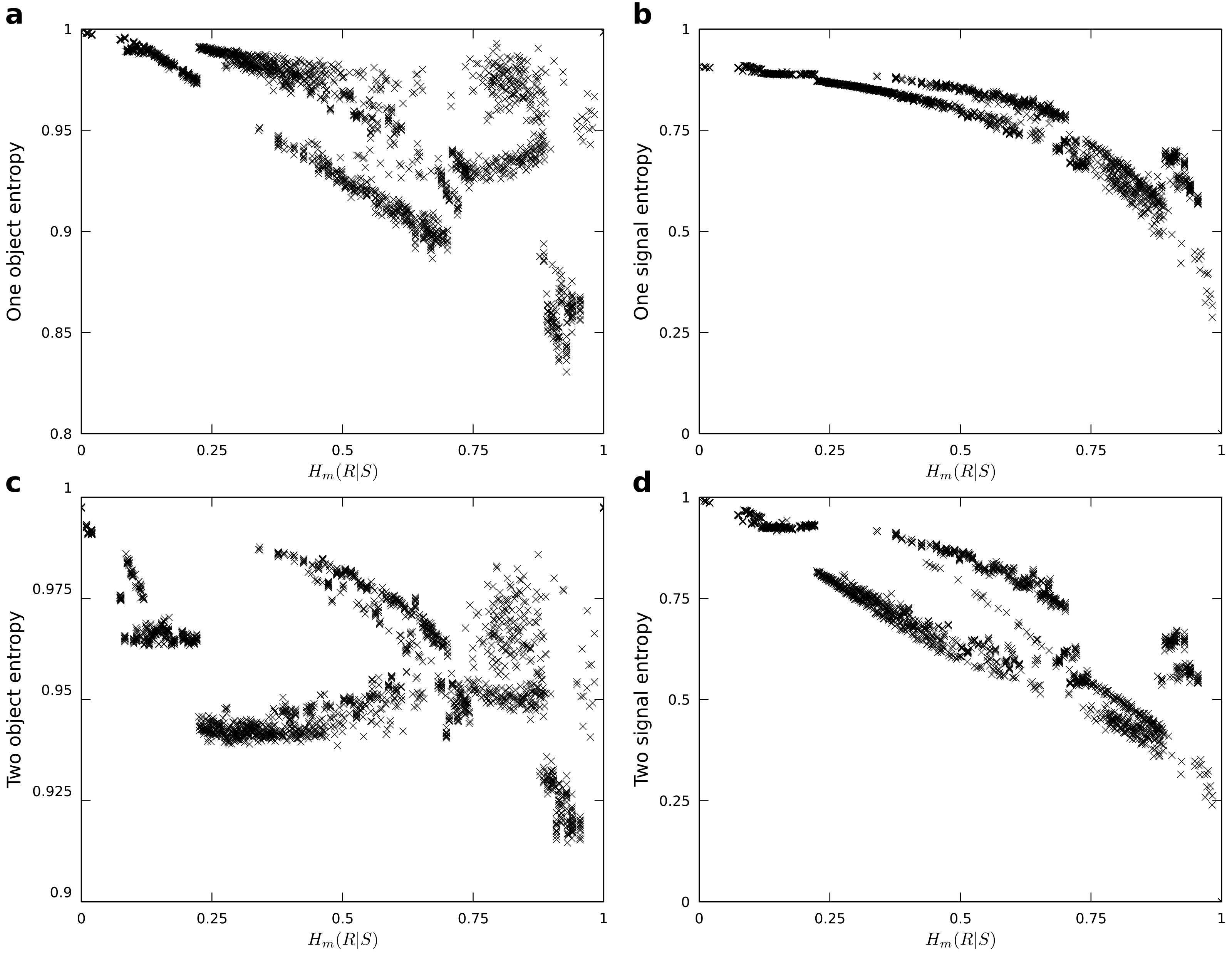}

          \caption{{\bf Complexity of codes as a random generative model along the Pareto front.} {\bf a} The entropy of objects as sampled by a random walker ($H_R$) over the language network is maximal at either end of the front and presents a minimum close to the star graph. {\bf b} The entropy of objects as sampled by a random walker ($H_S$) decreases rather smoothly along the Pareto front as we move from the one-to-one mapping to the star. {\bf c} The entropy of $2$-gram objects as sampled by a random walker ($H_{2R}$) presents less structure than $H_R$ and is still maximal at either extreme. {\bf d} The entropy of $2$-grams signals as sampled by a random walker ($H_{2S}$) also decreases as we move along the front, but in a less structured fashion. }

					\label{fig:10}
				\end{center}
			\end{figure*}

			By construction, $H_R$ must be maximum at both extremes of the front and non-trivial along it (figure \ref{fig:10}{\bf a}). In the one-to-one mapping, a same object is always sampled repeatedly, resulting in a reset of the random walk process as described in section \ref{sec:3.3}. Because the starting point is uniformly random, so must be the random walk and $H_R$ collapses to $1$. This results in a maximal entropy over signals as well (figure \ref{fig:10}{\bf b}). At the star graph, only one signal produces a valid sample of the code graph, and again this sample is uniform over objects (resulting in $H_R=1$, figure \ref{fig:10}{\bf a}) but this implies a maximally asymmetric sampling of words ($H_S=0$, figure \ref{fig:10}{\bf b}). Along the front, objects group up in clusters of different size, resulting in potentially greater biases towards some objects than others. This results in the possibility of a lower $H_R$, which is not always fulfilled. As in other cases, we see that a same point along the front hosts several different language networks with diverse $H_R$ values. The set of languages that produce a more remarkable structure is very close to the star graph (figure \ref{fig:10}{\bf a}). Overall, $H_R$ is large along the Pareto front as it was throughout the morphospace. The number of objects that a word links together determines how often that signal can be sampled through the random walker without reseting the process. This results in a smooth curve of decreasing entropy for $H_S$ (figure \ref{fig:10}{\bf b}). This suggests an explanation for the area of the morphospace with lowest $H_S$ in figure \ref{fig:05}{\bf b} near the star graph. 

			The entropy of $2$-gram objects also has to be maximal at both ends of the front (figure \ref{fig:10}{\bf c}). It remains largely unconstrained along the rest of the front, with little correlation and again large variability at a given point. The entropy of $2$-gram signals again decays to $0$ as the start graph is approached, but the decay is now less smooth and the range of values of $H_{2S}$ at a given point is larger.

		\subsection{Zipf, and other power laws}
			\label{app:02.04}

			Using the same methods as in section \ref{sec:3.4}, we computed the goodness of fit of word distribution to either Zipf or power-laws with arbitrary exponents. One of the caveats is that our languages across the morphospace are relatively small (n=400=m). While this is partly alleviated here (thanks to languages with up to $1\>000$ signals), these are nevertheless meager numbers. The results in this section can again mount evidence against the least-effort hypothesis as the origin of Zipf's distribution in human language, but this must be taken with extreme care given the computational shortages just mentioned. 

			\begin{figure*}
				\begin{center}
					\includegraphics[width=\textwidth]{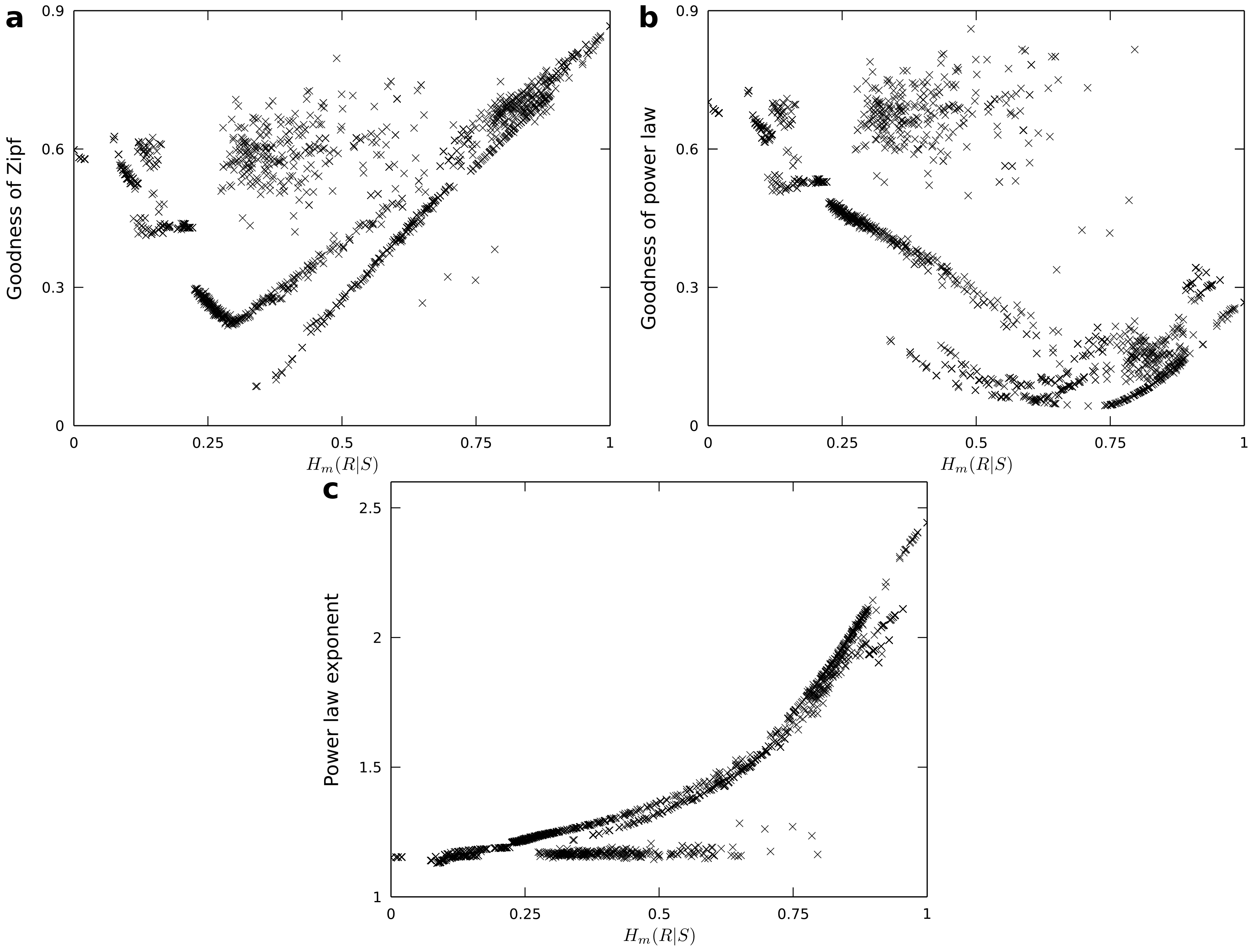}

          \caption{{\bf Power laws from the least-effort model along the Pareto front.} {\bf a} Goodness of fit of the word distribution from the toy, least-effort model to a Zipf law along the Pareto front. {\bf b} Goodness of fit of the word distribution from the model to an arbitrary power law along the Pareto front. {\bf c} Exponent obtained along the Pareto front when fitting the word distribution of the model to the arbitrary power law from panel {\bf b}. }

					\label{fig:11}
				\end{center}
			\end{figure*}

			Regarding goodness of fit to Zipf's law, along the Pareto front we find again a great variety of codes even within single points along the critical manifold (figure \ref{fig:11}{\bf a}). This indicates, as pointed out above and already anticipated in \cite{ProkopenkoPolani2010, SalgeProkopenko2013}, that least-effort alone would not be enough to enforce Zipf's distribution into word corpora -- at least not within this very limited toy model. There is a clear minimum of $KS$-score (i.e. maximum fitness to Zipf's distribution, figure \ref{fig:11}{\bf a}) around $\Omega_h \sim 0.3$ (hence $\Omega_s \sim 0.7$). This is close to, but not right at the value $\Omega_h = 1/2 = \Omega_s$ put forward in \cite{FortunyCorominas2013} for theoretical reasons. Also, the minimum KS-score ($\sim 0.1$) is larger than scores reached deeper inside the morphospace. According to this, the observation of Zipf's law in natural corpora would be evidence against the least-effort principles captured by the model. 

			Regarding the goodness of fit to arbitrary power laws (figure \ref{fig:11}{\bf b}), we find a more shallow minimum suggesting a broader region of interesting Pareto optimal languages. Looking at the exponents that come out of those fits (figure \ref{fig:11}{\bf c}), we find two branches as we move in the direction of increasing $H_m(R|S)$: i) a branch of roughly constant and low exponents close to $1$ (hence similar to Zipf's law), ii) a branch of exponents that increase monotonously with $H_m(R|S)$. It is difficult to asses which of these branches is yielding the lowest KS-score (best fit) in figure \ref{fig:11}{\bf b}. 

		\subsection{Code archetypes along the Pareto front}
			\label{app:02.05}

			Finally, as we did for the whole language morphospace, we analyzed possible archetypes clustering out of the measurements across the Pareto front. We moved into PC space and tried building $3$ and $5$ language archetypes using $k$-means clustering. For $k=3$ we found three relatively stable clusters: i) a few codes near the one-to-one graph, ii) a few others near the star network, and iii) all remaining codes along the front. However, the boundaries between the clusters changed notably after different initializations of the algorithm, sometimes leaving the third group almost without elements. With $k=5$, the clusters found were not stable at all, meaning that different instantiations of $k$-means would lump codes together in very different ways. Those clusters would also overlap when plotted along the Pareto front. 

			These results are very unlike the outcome for the whole morphospace. There, applying $k$-means several times with random initializations would consistently yield the same broad classes, which were clearly segregated across the morphospace with little overlap at their borders. Our inability to converge into well defined archetypes at the Pareto front is yet another indication of its huge diversity. We should also be careful about the previous clustering of Pareto optima within groups $I$ and $V$ (see section \ref{sec:4}). Fortunately, those classes reach deeper inside the morphospace and do not seem to depend so much on Pareto optimal solutions.

\end{document}